\documentclass{aa}  

\usepackage{graphicx}
\usepackage{txfonts}
\usepackage{hyperref}
\hypersetup{
     colorlinks = true,
     linkcolor = blue,
     anchorcolor = blue,
     citecolor = blue,
     filecolor = blue,
     urlcolor = blue
     }

\newcommand{\Rsun}{\ensuremath{\,{\rm R}_\odot}}                  % Solar radius symbol
\newcommand{\prism}{\texttt{PRISM}}     % small caps for prism
\newcommand{\Rearth}{\ensuremath{\,{\rm R}_\oplus}}                 % Earth radius symbol

\begin{document}

   \title{Simulations of starspot anomalies within TESS exoplanetary transit light curves -- II.}
\titlerunning{Simulations of starspot anomalies within TESS -- II}
   \subtitle{Forecasting the frequency of starspot anomalies appearing in TESS exoplanetary transit light curves}

   \author{J. Tregloan-Reed
          \inst{1}
           \and
          E. Unda-Sanzana\inst{2}
          }
\authorrunning{Tregloan-Reed 
\and 
Unda-Sanzana
}

   \institute{Instituto de Investigación en Astronomia y Ciencias Planetarias, Universidad de Atacama, Copiapó, Atacama, Chile \\
              \email{jeremy.tregloan-reed@uda.cl}
              \and
              Centro de Astronom\'{i}a (CITEVA), Universidad de Antofagasta,
              Avenida U. de Antofagasta 02800, Antofagasta, Chile
             }

   \date{Received Month dd,yyyy; accepted Month dd,yyyy}

% \abstract{}{}{}{}{} 
% 5 {} token are mandatory
 
  \abstract
  % context heading (optional)
  % {} leave it empty if necessary 
  {}
  % aims heading (mandatory)
   {We determine the starspot detection rate in exoplanetary transit light curves for M and K dwarf stars observed by the Transiting Exoplanet Survey Satellite (TESS) using various starspot filling factors and starspot distributions. }
  % methods heading (mandatory)
   {We used $3.6\times10^9$ simulations of planetary transits around spotted stars using the transit-starspot model \prism. The simulations cover a range of starspot filling factors using one of three distributions: uniform, polar-biased, and mid-latitude. After construction of the stellar disc and starspots, we checked the transit cord for starspots and examined the change in flux of each starspot to determine whether or not a starspot anomaly would be detected. The results were then compared to predicted planetary detections for TESS. }
  % results heading (mandatory)
   {The results show that for the case of a uniform starspot distribution, $64\pm9$ M dwarf and $23\pm4$ K dwarf transit light curves observed by TESS will contain a starspot anomaly. This reduces to $37\pm6$ M dwarf and $12\pm2$ K dwarf light curves for a polar-biased distribution and $47\pm7$ M dwarf and $21\pm4$ K dwarf light curves for a mid-latitude distribution.}
  % conclusions heading (optional), leave it empty if necessary 
   {Currently there are only 17 M dwarf and 10 K dwarf confirmed planetary systems from TESS, none of which are confirmed as showing starspot anomalies. All three starspot distributions can explain the current trend. However, with such a small sample, a firm conclusion can not be made at present. In the coming years when more TESS M and K dwarf exoplanetary systems have been detected and characterised, it will be possible to determine the dominant starspot distribution. }


   \maketitle
%
%-------------------------------------------------------------------

\section{Introduction}\label{Sec:1}

The NASA {\it Transiting Exoplanet Survey Satellite} (TESS) \citealt{TESS,TESS2,TESS3}) was developed to discover super-Earth(1.25\Rearth\ to 2\Rearth)-, sub-Neptune(2\Rearth\ to 4\Rearth)-, and Neptune(4\Rearth\ to 6\Rearth)-class planets transiting the brightest stars within the solar neighbourhood, the majority of which are K and M dwarf stars. \citet{Oneal2004} used studies of TiO
absorption bands to show that young K dwarfs can have a starspot filling factor of $20\%$\,--\,$40\%$. Similarly, it has been shown that young M dwarfs can have a starspot filling factor of $40\%\pm10\%$ \citep{Jackson2013}. While, the target stars of these two studies were young, the activity--age relationship of K and M dwarf stars is poorly known. However, the starspot coverage follows an inverse square root function with age \citep{Skumanich1972, Morris2020} towards solar levels ($<1$\%). This gives rise to the possibility that a larger population of TESS K and M dwarf targets may contain starspot anomalies within the observed light curves compared to G and F dwarf systems. 

The initial forecasts of predicted planetary yields from TESS indicated the potential to detect approximately 1700 planets from the 200,000 pre-selected stars \citep{Sullivan2015, Sullivan2015_erratum} observed with a cadence of 2\,min. However, since the launch of TESS, \citet{Barclay2018} revised these forecasts based on the performance of TESS and found that $1250\pm70$ will be discovered with the two-minute cadence observing mode. The revised forecast expects approximately 800 of the planets to be sub-Neptunes (2\Rearth to 4\Rearth), while, $\approx240$ super-Earth or smaller ($<$2\Rearth) planets are expected to be found \citep{Barclay2018}. It is expected that TESS will discover approximately 330 planets around M dwarf stars and approximately 100 around K dwarfs.

To distinguish between a rocky or gaseous planet requires a precise measurement of the planetary radius and mass. Planetary atmosphere studies using transmission spectroscopy \citep{Seager2000, Charbonneau2002} and/or multiband photometry \citep[e.g.][]{Sou2012, Sou2015, Southworth2018, Mancini2013a, Mancini2013b, Mancini2014, Chen2014, Jeremy2018} also require precise measurement of the planetary radius. Starspot anomalies can impact the shape of the light curve \citep{Silva2010} and if not correctly modelled can lead to biased measurements of the system parameters (e.g. planetary radius: \citealt{Nikolov2013}; limb-darkening coefficients: \citealt{Ballerini2012}; time of minimum light: \citealt{Sanchis2011a}), which would skew the photometric parameters of the light curve. However, if the perturbations are accounted for then precise measurement of the planetary radius will be possible, improving investigations into the atmosphere, structure, and evolution of such planets \citep{Fortney2007}.

Super-Earth and smaller planets detected by TESS orbiting M dwarf stars will be prime candidates for further in-depth atmospheric studies with the NASA {\it James Web Space Telescope} (JWST) searching for important molecular biomarker species (e.g. O$_3$, H$_2$O, CH$_4$ and CO$_2$). However, with a potential increase in the likelihood that these targets may contain starspot anomalies within the transit light curves, the  reliance of the community on dedicated transit-starspot models could increase. Nevertheless, many dedicated transit-starspot models have been developed by both the eclipsing binary star community (e.g. {\tt Wilson-Devinney} code: \citealt{wdcode,Wilson1979,Wilson1990,Wilson2008,Wilson2012}; {\tt PHEOBE}: \citealt{phoebecode,phoebecode2}) and the exoplanet community (e.g. \prism: \citealt{Jeremy2012,Jeremy2015,Jeremy2018}; {\tt SOAP-T}: \citealt{Oshagh2013}; {\tt spotrod}: \citealt{Beky2014}; {\tt KSint}: \citealt{Montalto2014}; {\tt ellc}: \citealt{Maxted2016}; {\tt StarSim}: \citealt{Herrero2016}; {\tt PyTranSpot}: \citealt{Juven2018}). While starspot anomalies in transit light curves are viewed as a nuisance, they open important new possibilities for understanding planetary systems. Indeed, they have the potential to allow determination of the rotation period of the host star \citep{Silva2008} and the sky-projected orbital obliquity \citep[e.g.][]{Jeremy2012, Jeremy2015, Sanchis2011a, Sanchis2011b, Sanchis2012, Sanchis2013, Mancini2013b, Mancini2014, Mohler2013} of the system directly from the light curve.

Currently, there are a total of 120\,\footnote{See \href{https://exoplanetarchive.ipac.caltech.edu/}{Exoplanet Archive} accessed 2021/03/01.} confirmed TESS exoplanets (e.g. Pi Mensae\,c: \citealt{Huang2018}: LHS 3884\,b; \citealt{Vanderspek2018}: HD 21749\,b; \citealt{Dragomir2019}: HD 202772A\,b; \citealt{Wang2019}), 25 of which are orbiting 17 M dwarf stars and 17 are orbiting 10 K dwarf stars.

The objective of the present study is to run simulations using various different starspot filling factors and distributions to forecast the number of TESS transit light curves that may harbour starspot anomalies. The outline of this paper is as follows. Section\,\ref{Sec:2} describes how the simulations were constructed and how the transit-starspot model \prism\  was used. Section\,\ref{Sec:3} presents the results for the different starspot distributions and filling factors, which are separated for K and M dwarf stars. In Sect.\,\ref{Sec:4} we discuss our results and provide the overall conclusions that we draw from the simulations.

%--------------------------------------------------------------------
\section{Simulation development}\label{Sec:2}

The simulations were developed to determine the likelihood that a TESS exoplanetary transit light curve would contain a starspot anomaly. Once this was completed for both K and M dwarf stars for a variety of planet radii, the results were extrapolated to the predicted TESS planetary yields given by \citet{Barclay2018}. By adopting this method, it was possible to allow each simulation to always result in a planetary transit, because the predicted planetary yields by \citet{Barclay2018} took into account false positives and non-detections.

The simulations in this work used the transit-starspot model {\it Planetary Retrospective Integrated Starspot Model} (\prism): \citep{Jeremy2012, Jeremy2015, Jeremy2018}. \prism\footnote{The latest version of \prism\ is available from \href{https://github.com/JTregloanReed/PRISM_GEMC}{GitHub}} is written in \textsc{idl}\footnote{For further details see \href{http://www.harrisgeospatial.com/ProductsandTechnology/Software/IDL.aspx}{\tt http://www.harrisgeospatial.com/Prod\\uctsandTechnology/Software/IDL.aspx}\,.} (Interactive Data Language) and models the stellar disc as a two-dimensional array via a pixelation approach to divide the star into many individual elements. Each element is then assigned an intensity value based on its radial distance from the centre of the stellar disc for the application of the quadratic limb-darkening law. The two-dimensional coordinates of the starspot(s) are then calculated and the intensity values of the elements within the starspot(s) are corrected for. The planet is then set to transit the star and the total received intensity is calculated based on which elements of the star are visible. By using the pixelation method, stellar features can be modelled by assigning individual intensities to each element \citep[see][]{Jeremy2012}.

The current version of \prism\ uses ten parameters to model the light curve, namely the planetary and stellar radii ratio\,($k$), the sum of the fractional ($r_\mathrm{p} = R_\mathrm{p} / a$ and $r_\mathrm{*} = R_\mathrm{*} / a$), planetary, and stellar radii\,($r_\mathrm{p} + r_\mathrm{*}$), the linear and quadratic coefficients of the quadratic limb-darkening law\,($u_1$ and $u_2$), orbital inclination\,($i$), the time of minimum light\,($T_0$), orbital eccentricity\,($e$), the argument of periastron\,($\omega$), third light ($l_3$), and the out-of-transit detrending polynomial coefficients ($c_0,\,c_1\,...\,c_\mathrm{n}$) \citep{Jeremy2012, Jeremy2015, Jeremy2018}, combined with four additional parameters for each starspot: longitude of the spot's centre\,($\theta$), co-latitude of the spot's centre\,($\phi$), angular size\,($r_\mathrm{spot}$), and contrast\,($\rho_\mathrm{spot})$ \citep{Jeremy2012}.

For the simulation presented in this work, \prism\ was used to model the stellar disc, the starspot(s), and the transit cord. \prism\ has undergone various upgrades and optimisations since its initial release. A recent change optimised the way \prism\ scans the elements of the stellar disc that lay within the starspot. This reduces the time required to build the stellar disc array and populate it with starspots by more than an order of magnitude\footnote{Based on a 3.6\,Ghz Hex core processor.}. However, to simplify the coding changes, $\phi$ was altered to represent the stellar latitude instead of the co-latitude.

%--------------------------------------------------------------------
\subsection{Simulation parameters}\label{Sec:2.1}

A study using high-resolution Doppler imaging of the two M dwarfs, GJ\,791.2\,A and LP\,944-20, indicated a large starspot filling factor at high latitudes \citep{Barnes_J2015}. The study determined an overall starspot filling factor of a few percent for the two stars, but at high latitudes the starspot filling factor was determined to be around $80\%$ \citep{Barnes_J2015}. On the other hand, for known K dwarf planetary systems that exhibit starspot anomalies, the starspots appear in the mid-latitude range (e.g. HAT-P-11, \citealt{Sanchis2011b}), agreeing with predictive dynamo models \citep[e.g.][]{Moss2011}. Therefore, the simulations used three starspot distributions: uniform, mid-latitude, and polar-biased.

The first parameter that directly influences whether or not a planet occults a starspot is the starspot filling factor. A planet is more likely to occult a starspot on a host star with a higher starspot filling factor. It has been shown that for young K and M dwarfs the starspot filling factor is $30\%\pm10\%$ \citep{Oneal2004} and $40\%\pm10\%$ \citep{Jackson2013} respectively. Because the activity--age relationship of K and M dwarfs has not been constrained, the simulations were conducted for different starspot filling factors (simulating different ages), starting at those given in the literature for young stars. The starspot filling factor was then reduced in logarithmic steps (i.e. for the M dwarf simulations: $40\%\pm10\%$, $20\%\pm5\%$, $4.0\%\pm1.0\%$, $2.0\%\pm0.5\%$, $0.4\%\pm0.1\%$\%, and $0.2\%\pm0.05\%$) to represent the inverse square root decay of starspot coverage with age \citep{Skumanich1972, Morris2020}, whilst reproducing the small covering fraction observed on the Sun ($<1\%$) and older K-dwarfs.

The second important parameter that affects the likelihood that a planet occults a starspot is $k$. For fixed $R_\mathrm{*}$, a large planet is more likely to occult a starspot than a smaller planet because the the transit cord covers a larger proportion of the stellar disc. Similarly, for fixed $R_\mathrm{p}$, a larger star will reduce the likelihood that a planet will occult a starspot because of the reduction in the area of the transit cord. Previous studies constrained the relationship between the smallest detectable change in flux of a starspot ($\Delta F_{spot}$) and the change in flux ($\Delta F_p$) at the time of minimum light of a planetary transit observed by TESS \citep{Jeremy2019}. The results showed that the smallest detectable $\Delta F_{spot}$ was when $k=0.082\pm0.004$.

The orbital inclination plays an important role in determining whether or not a planet occults a starspot. When assuming a uniform starspot distribution, $i=90^\circ$ will give the greatest likelihood of a starspot crossing event because it maximises the area of the stellar disc covered by the transit cord. However, for a polar-biased starspot distribution, a value of $i$ that gives a near grazing or grazing transit will have the greatest likelihood\footnote{Assuming spin orbit alignment (see Section\,\ref{Sec:2.2.1}). } of a starspot crossing event. For these simulations we instead chose to use the impact parameter $b$ instead of $i$. This is because $b$ encompasses $i$, the semi-major axis $a$; the orbital eccentricity $e;$ and the argument of periastron $\omega$ into a single parameter. \citep[e.g.][]{Winn2010Book}:  

\begin{equation} \label{Eq.1}
b = \frac{a \cos i}{R_\mathrm{*}} \left(\frac{1-e^2}{1+e\sin\omega}\right) \ ,
\end{equation}

\noindent all of which influence which stellar latitude the transit cord intersects. 

The orbital period $P$ directly correlates with $a$ via Kepler's third law, therefore using $b$ takes into account the various permutations between $i$, $a$, $P$, $e,$ and $\omega$. Using $b$ also streamlines the simulations in terms of simplicity. If $i$ was used then an additional layer of calculations would be needed to guarantee a transit (by calculating the limit in $i$), while by using $b$ this limit is simply set at $\pm(R_\mathrm{*} + R_\mathrm{p})$.

The final set of parameters that influence whether or not a starspot anomaly is detected belong to the starspots themselves ($\theta$, $\phi$, $ r_\mathrm{spot}$, and $\rho_\mathrm{spot}$). For the two starspot distributions we consider in this study, the position of a starspot on the stellar disc becomes inconsequential. Namely, in the case of a uniform starspot distribution there may be many starspots on the stellar disc at various different latitudes, and therefore $b,$ along with the starspot filling factor and $k$ will primarily determine whether or not the starspot is occulted. On the other hand, in the case of a polar-biased starspot distribution, the latitude of the starspot will be constrained to high latitudes and hence only transit cords that cross high latitudes will detect the starspot. However, the size and contrast of the starspot determine if an anomaly is detected, unlike the other parameters which determine if a starspot crossing event takes place. The size and contrast of the starspot affect the size of the starspot anomaly. This means that a starspot-crossing event can take place, but if the starspot is too small and/or has a low contrast (e.g. $\rho_\mathrm{spot} > 0.99$) the event goes undetected \citep[e.g.][]{Jeremy2019}.

%The final set of parameters which influences if a starspot anomaly is detected, belong to the starspots themselves. For a uniform starspot distribution, the position of a starspot on the stellar disc becomes inconsequential considering that there may be many starspots on the stellar disc at various different latitudes, and that, $b$ along with the starspot filling factor and the width of the transit cord ($k$), will primarily determine if the starspot is occulted. This is also true for a polar biased starspot distribution as the latitude of the starspot will be constrained to high latitudes. Hence, only transit cords that cross high latitudes will detect the starspot. However, the contrast and size of the starspot do determine if an anomaly is detected. Unlike the other parameters which determine if a starspot crossing event takes place. The size and contrast of the starspot affect the size of the starspot anomaly. This allows a starspot crossing event to take place, but, if the starspot is to small and/or has a low contrast (e.g., $\rho_\mathrm{spot} > 0.99$) the event goes undetected \citep[e.g.,][]{Jeremy2019}.

%--------------------------------------------------------------------
\subsection{Simulation parameter space}\label{Sec:2.2}

\begin{figure*} \centering \includegraphics[width=0.49\textwidth,angle=0]{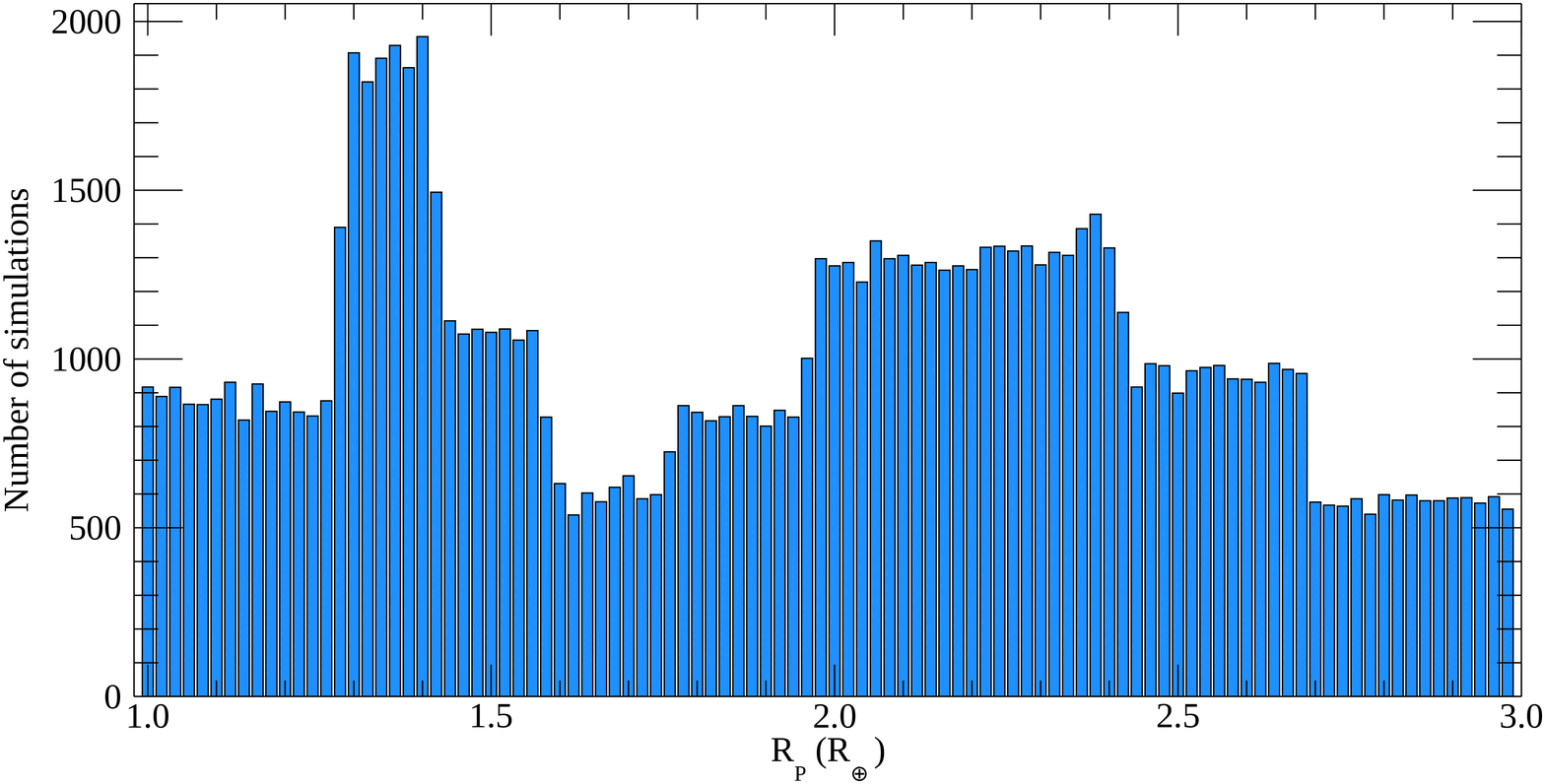} 
\includegraphics[width=0.49\textwidth,angle=0]{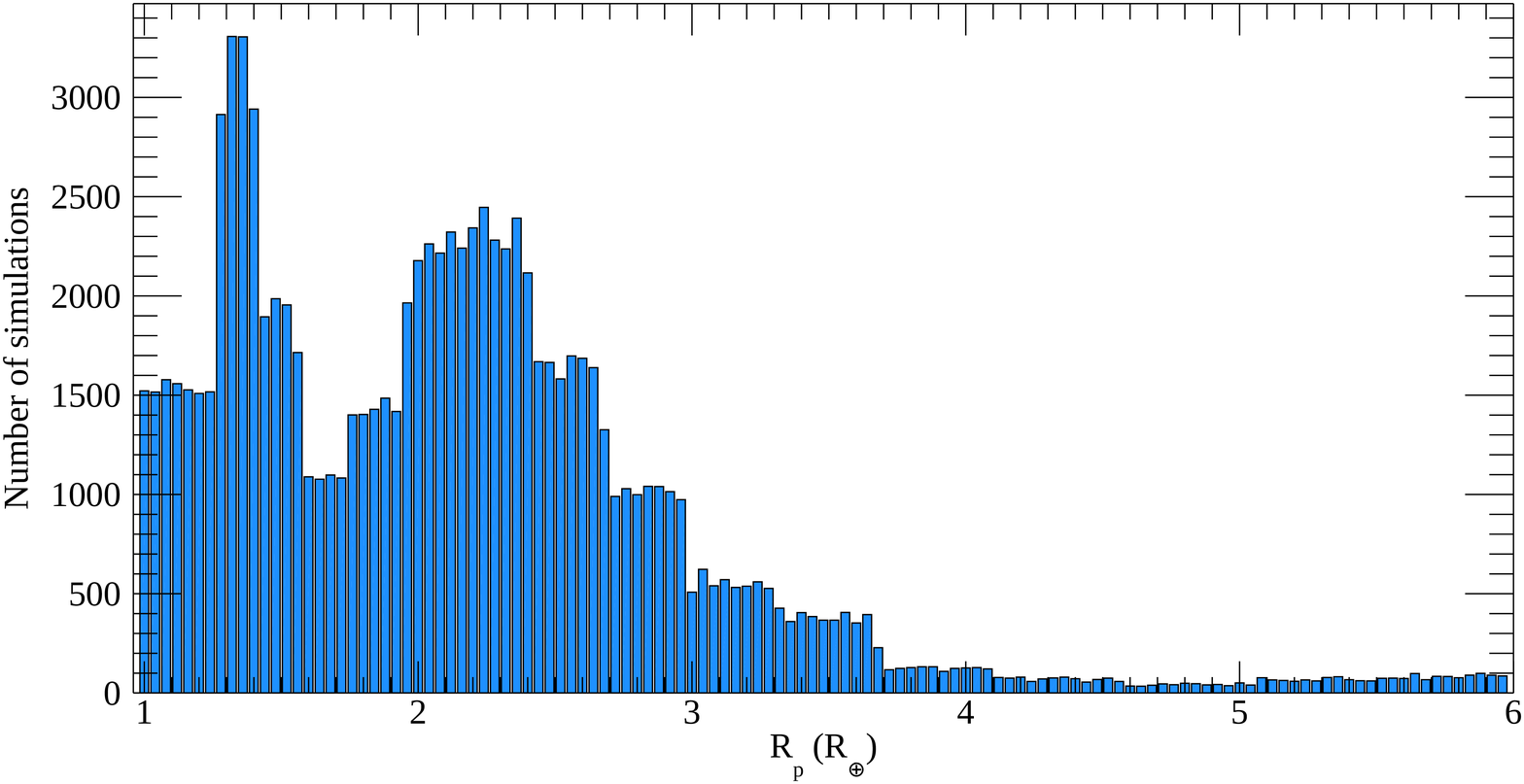}
\caption{\label{fig:1a} Frequency of $R_\mathrm{p}$ used in the two simulation runs, which shows the shape of the planetary radius distribution used in the simulations. {\it Left} from simulation run \#\,{\it229} using a polar-biased starspot distribution of a M dwarf host star with a starspot filling factor of $4.0\%\pm1.0\%$. {\it Right} from simulation run \#\,{\it807} using a mid-latitude starspot distribution of a K dwarf host star with a starspot filling factor of $0.3\%\pm0.1\%$.} \end{figure*}

Each simulation randomly selected values for the parameters described in Section\,\ref{Sec:2.1}. Each selection either drew from a uniform distribution (e.g. impact parameter) or from a Gaussian distribution (e.g. starspot filling factor). However, the planetary radius parameter was drawn from a distribution which was built to simulate the power-law rise in occurrence rates from 5.7\,\Rearth\ to 2\,\Rearth\ \citep[see][]{Petigura2013}, the astrophysically driven valley near 2\,\Rearth\ due to photo-evaporation \citep[see][]{Fulton2017}, and the detection efficiency bias which skews the distribution of detected planets to larger planet radii. Figure\,\ref{fig:1a} shows the frequency of selected $R_\mathrm{p}$ values used in two simulation runs (one for an M-dwarf and the second for a K-dwarf), which highlights the distribution of values from which $R_\mathrm{p}$ was drawn.

Starspot filling factors were determined for young K an M dwarfs, but their activity--age relationship has not been constrained. As a consequence, we ran simulations for different distributions of the starspot filling factor. The starting distributions were set at the values determined for young K and M dwarfs at $30\%\pm10\%$ \citep{Oneal2004} and $40\%\pm10\%$ \citep{Jackson2013}, respectively. Each sequential distribution was logarithmically reduced to values of $<1\%$ to simulate solar filling factors and older K dwarfs. Additionally, a second distribution range was used, which was set at half the filling factors of the logarithmically reduced values (e.g. $4.0\%\pm1.0\%$ and $2.0\%\pm0.5\%$). Table\,\ref{Tab.1} lists the 12 distributions of the starspot filling factor for both K and M dwarf simulations.  

\begin{table} \centering
\caption{\label{Tab.1} Gaussian distributions of the starspot filling factors for K and M dwarfs used in this work.}
\setlength{\tabcolsep}{6pt} \vspace{-5pt}
\begin{tabular}{cc} 
\hline\hline
 M dwarf star & K dwarf star \\   
\hline
$40\%\pm10\%$ & $30\%\pm10\%$ \\
$20\%\pm5\%$ & $15\%\pm5\%$ \\
$4.0\%\pm1.0\%$ & $3.0\%\pm1.0\%$ \\
$2.0\%\pm0.5\%$ & $1.5\%\pm0.5\%$ \\
$0.4\%\pm0.1\%$ & $0.3\%\pm0.1\%$ \\
$0.2\%\pm0.05\%$ & $0.15\%\pm0.05\%$ \\
\hline \end{tabular} 
\end{table}

To allow a direct extrapolation from the results from this work using the predicted TESS planetary yields, we used the same effective temperature ($T_\mathrm{eff}$) range used by \citet{Sullivan2015,Barclay2018}. These latter authors used $T_\mathrm{eff}$ from 3200\,K (M4V) to 3700\,K (M1V) and 4100\,K (K5V) to 5000\,K (K1V). A lower limit of 3200\,K was used for the M dwarf stars because M dwarfs cooler than 3200\,K are too faint for transit detection in the TESS passband \citep{Sullivan2015}. However, for the K dwarf simulations we extend the range to include K9V dwarf stars with $T_\mathrm{eff} = 3900$\,K \citep{Habets1981}. 

$R_\mathrm{*}$ in units of \Rsun\ was then calculated using a third-order polynomial derived from non-evolved K and M stars within the temperature range 3200\,K to 5500\,K using interferometric measurements of M\,and\,K dwarfs \citep[Eq.\,6][]{Boyajian2012}: 

\begin{equation}\label{Eq.2}
\begin{split}
R\left(\Rsun\right) = & -\left(10.8828\pm0.1355\right) + \left(7.18727\pm0.09468\right)\times10^{-3}\,T_\mathrm{eff} \\
& -\left(1.50957\pm0.02155\right)\times10^{-6}\,T_\mathrm{eff}^2 \\
& + \left(1.07572\pm0.01599\right)\times10^{-10}\,T_\mathrm{eff}^3\ .
\end{split}
\end{equation}

\noindent The distributional range of $R_\mathrm{*}$ derived from the distributional range of $T_\mathrm{eff}$ using Eq.\,\ref{Eq.2} are given in Table\,\ref{Tab.2}. When constructing the distribution of $R_\mathrm{*}$ from the distribution of $T_\mathrm{eff}$ we ignore the coefficient uncertainties in Eq.\,\ref{Eq.2}. This is an acceptable simplification because we are constructing a distribution that incorporates the coefficient uncertainties. The coefficient uncertainties allow different values of $T_\mathrm{eff}$ to be mapped to the same value of $R_\mathrm{*}$. Therefore, randomly picking a value of $R_\mathrm{*}$ automatically selects all the different values of $T_\mathrm{eff}$ which map to $R_\mathrm{*}$ via Eq.\,\ref{Eq.2}. Because only $R_\mathrm{*}$ affects the likelihood of a planet occulting a starspot,  which of the mapped $T_\mathrm{eff}$ was used to calculate $R_\mathrm{*}$  becomes irrelevant.

\begin{table} \centering
\caption{\label{Tab.2} Extrapolated values of $R_\mathrm{*}$ using $T_\mathrm{eff}$ from Eq.\,\ref{Eq.2} for the spectral type boundaries of K and M dwarfs used in this work.}
\setlength{\tabcolsep}{12pt} \vspace{-5pt}
\begin{tabular}{lcc} 
\hline\hline
 Spectral  & Range of $T_\mathrm{eff}$ & Range of $R_\mathrm{*}$ \\   
Type  & ($K$) & (\Rsun) \\
\hline
M dwarf & 3200--3700  & 0.155--0.493 \\ 
K dwarf & 3900--5000 & 0.568--0.761 \\ 
\hline \end{tabular} 
\end{table}

For the simulations involving M dwarf stars, $T_\mathrm{eff}$ was selected from a uniform distribution between 3200\,K and 3700\,K and the resulting $T_\mathrm{eff}$ was used with Eq.\,\ref{Eq.2} to determine $R_\mathrm{*}$. This was the equivalent of selecting $R_\mathrm{*}$ from a uniform distribution between 0.155\Rsun\ and 0.493\Rsun. For the K dwarf stars $T_\mathrm{eff}$ was selected from a uniform distribution between 3900\,K and 5000\,K, giving the equivalent of selecting $R_\mathrm{*}$ from a uniform distribution between 0.568\Rsun\ and 0.761\Rsun. 

It follows from Section\,\ref{Sec:2.1} that to ascertain a value for $k$, values for $R_\mathrm{*}$ and $R_\mathrm{p}$ need to be selected. In the M dwarf simulation, $R_\mathrm{p}$ was selected from the constructed distribution based on the results from \citet{Petigura2013,Fulton2017} between 1\Rearth\ and 3\Rearth. The lower limit of 1\Rearth was selected as this produced a transit depth of $0.35\%$ for the M4V dwarf star and $0.035\%$ for the M1V dwarf star. TESS has a photometric sensitivity of 350\,ppm for an $V = 12$ star \citep{Barclay2018}, with over 800 exoplanet candidate detections. Therefore, the two transit depths should be detectable by TESS. An upper limit for $R_\mathrm{p}$ was set at 3\Rearth. \citet{Dressing2013, Dressing2015, Morton2014} determined from population studies of Kepler and K2 data that close-in ($<$\,50\,d), large ($>$\,3\Rearth) planets orbiting M dwarf stars have a low occurrence rate. However, there are a few exceptions (e.g.  NGTS-1\,b; \citealt{Bayliss2018}: Kepler-45\,b; \citealt{Johnson2012}: HATS-6\,b; \citealt{Hartman2015}).

For the simulations that used a K dwarf star, $R_\mathrm{p}$ was selected from the constructed distribution based on the results from \citet{Petigura2013,Fulton2017} between 1\Rearth\ and 6\Rearth. The upper limit of 6\Rearth \ was selected to be in line with the TESS mission to find small planets $\le$\,6\Rearth. For an $V = 10.5$ star, TESS can achieve a photometric precision of 200\,ppm \citep{Barclay2018}. A 1\Rearth\ planet transiting a K9V dwarf will produce a transit depth of $0.026\%$ (260\,ppm) while, if transiting a K1V dwarf, it will produce a transit depth of $0.014\%$ (145\,ppm), which TESS can detect for stars brighter than $V = 9.7$. We calculated that a 1.17\Rearth\ planet transiting a K1V dwarf would produce a transit depth of 200\,ppm. We therefore find it acceptable to select $R_p$ between 1\Rearth\ and 6\Rearth\ for all K dwarfs in these simulations. 

For each simulation, $b$ was selected from a uniform distribution between $-\left(R_\mathrm{*} + R_\mathrm{p}\right)$ and $\left(R_\mathrm{*} + R_\mathrm{p}\right)$. In addition to the advantage of using $b$ over $i$ given in Section\,\ref{Sec:2.1}, setting the range of $b$ for the entire stellar disc (see Section\,\ref{Sec:2.2.1}) guaranties a transit for each simulation. The results from these simulations are applied to the predicted TESS planetary yields below, where we also take into account the likelihood of false positives and non-detections.

The four starspot parameters ($\theta$, $\phi$, $ r_\mathrm{spot}$, and $\rho_\mathrm{spot}$) were selected from uniform distributions. Because only half the stellar surface is visible, the location of the starspot was kept within the visible stellar disc: $-90^\circ\le \theta \le +90^\circ$ and $0^\circ\le \phi \le180^\circ$. Here, $ r_\mathrm{spot}$ was selected between 0$^\circ$ and 20$^\circ$, in line with previous transit-starspot analyses using \prism\ \citep[e.g.][]{Jeremy2012,Jeremy2015,Mancini2017,Raetz2019}.

\citet{Silva2003} gives an equation to find $\rho_\mathrm{spot}$ using the stellar effective temperature ($T_\mathrm{eff}$), the temperature of the starspot ($T_\mathrm{spot}$), and the frequency of the observation ($\nu$):

\begin{equation} \label{eq.3}
 \rho_\mathrm{spot} = \frac{\exp\left(h\nu / k_\mathrm{B} T_\mathrm{eff}\right) - 1}{\exp\left(h\nu / k_\mathrm{B} T_\mathrm{spot}\right) - 1} \ ,
\end{equation}

\noindent where $h$ is Planck's constant and $k_B$ is Boltzmann constant. 

Using typical values of $T_\mathrm{spot}$ for M4V, M1V, and K5V stars observed by TESS, \citet{Jeremy2019} found $\rho_\mathrm{spot}$ to range from 0.53 to 0.93. Therefore, $\rho_\mathrm{spot}$ was selected from a range of 0.5 to 1.0 for the simulations. While $\rho_\mathrm{spot}$ does not determine whether or not a starspot intercepts with the transit cord, it does determine if the size of the starspot anomaly is detectable over the noise. A previous set of simulations were used to determine whether or not starspot anomalies could be detected in TESS planetary transit light curves, and if so, the minimum size that could be detected \citep{Jeremy2019}. The simulations used an array of parameters including but not limited to $R_\mathrm{*}$, $T_\mathrm{eff}$, $R_\mathrm{p}$, $ r_\mathrm{spot}$, $ T_\mathrm{spot}$, and $\rho_\mathrm{spot}$. The simulations indicate that the smallest detectable change in flux of a starspot in TESS light curves was when $k=0.082\pm0.004$ \citep{Jeremy2019}. To incorporate the results from \citet{Jeremy2019}, the change in flux of each starspot within the transit cord was calculated and compared to $k$ to determine whether or not the starspot was detectable using the following relation \citep[Eq.\,9][]{Jeremy2019}:

\begin{equation}\label{eq.4} 
\begin{split}
\Delta F_\mathrm{spot} = \left(0.00028\pm0.00001\right)t^2 & \\ 
- \left(0.00121\pm0.00001\right)t & \\ 
+ \left(0.00147\pm0.00002\right) \ ,
\end{split} 
\end{equation}

\noindent where $t$ is the logarithmic parametrisation of $k^2$:

\begin{equation}\label{eq.6}
 t = \log_{10}\left(\frac{1}{k^2}\right) \ .
\end{equation}

Additionally, one of the main factors that determine whether or not a starspot anomaly is detected in a transit light curve is the ratio between $\Delta F_\mathrm{spot}$ and the precision of the light curve. For TESS exoplanet light curves, $\Delta F_\mathrm{spot}$ needs to be at least twice the rms scatter of the light curve to fully constrain the starspot properties. A $\Delta F_\mathrm{spot}$ that is 1.5 times the rms scatter will allow the starspot to be detected, but it will not allow for the starspot parameters to be fully constrained \citep{Jeremy2019}.

We used the predicted planetary yields from \citet{Barclay2018} to create a distribution of light-curve rms scatters based on the TESS magnitude distribution of planet-hosting stars. For each simulation, the rms scatter of the light curve was selected from this distribution and the calculated $\Delta F_\mathrm{spot}$ values of starspots within the transit cord were compared to this value. If at least one $\Delta F_\mathrm{spot}$ was 1.5 times the rms scatter or greater, the light curve was deemed to contain a detectable starspot anomaly.

%--------------------------------------------------------------------
\subsubsection{Polar-biased starspot distribution}\label{Sec:2.2.1}

\begin{table*} \centering
\caption{\label{Tab.3} Fourteen planetary systems with measured $\lambda$ and/or $\psi$ with a stellar $T_\mathrm{eff}\le5000$\,K. All multiple measurements for the same object are included together with the associated weighted mean. The values in bold were used to determine the weighted mean of $\lambda$ from the 13 planetary systems with a measured $\lambda$. Obtained from the 2019/12/30 version of the TEPCat catalogue \citep{Me11mn}.}
\setlength{\tabcolsep}{8pt} \vspace{-5pt}
\begin{tabular}{lcccr} 
\hline\hline
System  & $T_\mathrm{eff}$ & $\lambda$ & $\psi$ & Reference \\   
              & ($K$)                     & ($^\circ$)   &  ($^\circ$)     & \\
\hline
GJ\,0436   &     $3416\pm54$     &    $\mathbf{72^{+33}_{-24}} $ & $80^{+21}_{-18}$  &\citep{Bourrier2018} \\
\hline
HAT-P-11   &    $4780\pm50$  &    $103^{+26}_{-10}$   &   & \citep{Winn2010b}    \\ 
HAT-P-11   &    $4780\pm50$  &    $103^{+22}_{-18}$   &   & \citep{Hirano2010}    \\ 
HAT-P-11   &    $4780\pm50$  &    $106^{+15}_{-12}$   & $106^{+15}_{-11}$  &  Solution 1\citep{Sanchis2011b}    \\ 
HAT-P-11   &    $4780\pm50$  &    $121^{+24}_{-21}$   &  $97^{+8}_{-4}$  &  Solution 2\citep{Sanchis2011b}    \\ 
HAT-P-11   &      &  $\mathbf{106.6\pm8.3}$   &    &      \\ 
\hline
HAT-P-12    &   $4665\pm45$  & $\mathbf{-54^{+41}_{-13}}$ & & \citep{Mancini2018} \\
\hline
HAT-P-18   &    $4870 \pm50$  &   $\mathbf{132\pm15}$     &           &     \citep{Esposito2014} \\   
\hline
HAT-P-20   &    $4595\pm45$ &    $\mathbf{-8.0\pm 6.9}$  &  $36^{+10}_{-12}$ &     \citep{Esposito2017}         \\
\hline
Qatar-1    &    $4910\pm100$ &   $\mathbf{-8.4\pm7.1}$   &              & \citep{Covino2013} \\  
\hline
Qatar-2    &    $4645\pm50$ &   $4.3\pm4.5$   &              & \citep{Mancini2014} \\   
Qatar-2    &    $4645\pm50$ &   $0.0\pm8.0$   &               &\citep{Mocnik2017} \\   
Qatar-2    &    $4645\pm50$ &   $0.0\pm10.0$   &               &\citep{Dai2017} \\   
Qatar-2    &    $4645\pm50$ &   $15\pm20$   &     $0^{+43}_{-0}$ &         \citep{Esposito2017} \\   
Qatar-2   &      &  $\mathbf{3.2\pm3.6}$   &    &      \\ 
\hline
WASP-11  &    $4900\pm65$ &    $\mathbf{7\pm5}$ &                 &     \citep{Mancini2015} \\
\hline
WASP-43   &    $4520\pm120$      &    $\mathbf{3.5\pm6.8}$ &  $0^{+20}_{-0}$ &      \citep{Esposito2017} \\
\hline
WASP-52   &    $5000\pm100$      &    $24^{+17}_{-9}$ &      &  \citep{Hebrard2013} \\
WASP-52   &    $5000\pm100$      &    $3.8\pm8.4$ &  $20\pm50$    &  \citep{Mancini2017} \\
WASP-52   &                      &    $\mathbf{9.7\pm7.1}$ &      &  \\
\hline
WASP-53   &    $4950\pm60$      &    $\mathbf{-1.0\pm12}$ &      &  \citep{Triaud2017} \\
\hline
WASP-69   &    $4700\pm50$      &    $\mathbf{0.4^{+2.0}_{-1.9}}$ &       & \citep{Casasayas-Barris2017} \\
\hline
WASP-80   &    $4145\pm100$      &    $0\pm20$ &        &\citep{Triaud2013} \\
WASP-80   &    $4145\pm100$      &    $-14\pm14$ &      &  \citep{Triaud2015} \\
WASP-80  &      &  $\mathbf{-9.4\pm11.5}$   &    &      \\ 
\hline
WASP-107\tablefootmark{a}   &    $4430\pm120$      &     &   $90\pm50$     &\citep{Dai2017} \\
\hline \end{tabular} \\
\tablefoot{
\tablefoottext{a}{\citet{Dai2017} only gives a measurement of $\psi$ for WASP-107. Therefore, WASP-107 was not used in calculating the weighted mean of $\lambda$.}
}
\end{table*}

When dealing with a polar-biased starspot distribution, an extra parameter needs to be considered, namely $\lambda$ (or $\psi$). In the case of a uniform distribution, the orientation of the stellar sphere is unimportant. This is due to a starspot having equal likelihood of appearing on the stellar equator or at a pole. Therefore the likelihood that a starspot will be occulted by a planet with $i=90^\circ$ is the same for both equatorial and polar transits. However, this is not the case for polar-biased starspot distributions. 

For an aligned system (i.e. $\psi \approx 0^\circ$ and $\lambda \approx 0^\circ$) a transiting planet with $i=90^\circ$ will not occult a polar starspot. However, if the stellar rotation axis lies parallel to the transit cord (i.e. $\lambda = \pm90^\circ$), a transiting planet with $i=90^\circ$ will occult a polar starspot at either ingress or egress.

At present, $\lambda$ has been measured for 147 planetary systems\footnote{Obtained from the 2021/03/08 version of TEPCat \citep{Me11mn}. \href{http://www.astro.keele.ac.uk/jkt/tepcat/obliquity.html}{\texttt{http://www.astro.keele.ac.uk/jkt/tepcat/obliquity.html}}.}, with stellar $T_\mathrm{eff}$ spanning 3400\,K to 9600\,K. It has been shown that hot stars ($T_\mathrm{eff}>6250$\,K) are more likely to harbour misaligned systems \citep[see][]{Winn2010c, Albrecht2012, Jeremy2015}. At 6250\,K, the mass of the convective envelop in a main sequence star drastically reduces \citep{Pinsonneault2001}. \citet{Winn2010c} discusses the finding that a reduction in the mass of the convective envelope reduces the magnetic fields produced by the convective envelope, and by doing so reduces the tidal dissipation. As a consequence a longer timescale is required for a planetary system to realign for hot stars ($T_\mathrm{eff}>6250$\,K). 

The highest $T_\mathrm{eff}$ used in the simulations is 5000\,K (see Section\,\ref{Sec:2.2}). Therefore, we assume that a larger proportion of planets will be aligned. To confirm this, we calculated the weighted mean of $\lambda$ from the 13 planetary systems (see Table\,\ref{Tab.3}) with a measured $\lambda$ and whose host stars $T_\mathrm{eff}\le5000$\,K. Where multiple measurements exist for the same object we adopt the weighted mean of these values in the use of the final weighted mean calculation (see Table\,\ref{Tab.3}). We find $\bar{\lambda} = 7.3^\circ\pm1.4^\circ$, which agrees with the findings of previous studies that the majority of cool stars have aligned systems \citep{Winn2010c, Albrecht2012, Jeremy2015}. When we set the boundary between alignment and misalignment at $\pm\,10^\circ$ \citep[e.g.][]{Winn2010c}, $69\%$ of systems show alignment. Using this result, we constructed a Gaussian distribution of $0^\circ\pm10^\circ$ to randomly draw $\lambda$ values for the simulations using a polar-biased starspot distribution, which leads to $68.2\%$ of randomly generated $\lambda$ values ranging between $-10.0^\circ$ and $10.0^\circ$.

To confirm if a starspot can be detected, we need to discern whether or not the latitude of the starspot intercepts with the transit cord. It is harder to constrain $\psi$ than $\lambda$ because of the difficulty in measuring the stellar inclination. This is borne out when examining the planetary systems with measured $\lambda$ given in Table\,\ref{Tab.3}. For the sample used in this work only 7 measurements of $\psi$ have been obtained compared to 21 measurements for $\lambda$. A similar trend extends to all 135 planetary systems with a measured $\lambda$. Just 34 measurements of $\psi$ have been accomplished compared to 210 measurements for $\lambda$. Because of a small sample of $\psi$ measurements (seven), the simulations using a polar-biased starspot distribution only used $\lambda$ to determine spin--orbit alignment. Therefore, the position of the starspots was constrained to the visible surface of the star. This is an acceptable simplification because any starspots that lie on the unseen surface (i.e. $90^\circ\le\theta\le270^\circ$; where $270^\circ \equiv -90^\circ$) will become visible as the star rotates.

In the simulations where a polar-biased starspot distribution was used, the latitude of each starspot was selected from one of two Gaussian distributions of $0^\circ$ to $30^\circ$ for the northern pole and $150^\circ$ to $180^\circ$ for the southern pole. The distributions were centred at the poles (i.e. $0^\circ$ and $180^\circ$) with the $30^\circ$ limits set at 2-$\sigma$ deviations, allowing $95.4\%$ of the generated starspots to lie within $30^\circ$ of each pole. This restriction was used to emulate the starspot distributions found on GJ\,791.2\,A and LP\,944-20 where a high starspot filling factor was found at high latitudes, while a starspot filling factor of a few percent was found at lower latitudes \citep{Barnes_J2015}.

%--------------------------------------------------------------------
\subsubsection{Mid-latitude starspot distribution}\label{Sec:2.2.2}

For mid-latitude starspot distributions, $\lambda$ (and $\psi$) play an important role in the orientation of the stellar disc, which influences whether or not a starspot is detected for a given $i$. As a consequence, for the mid-latitude simulations, values of $\lambda$ were drawn from the same Gaussian distribution that was used for the polar-biased simulations, in that 1-$\sigma$ of randomly generated $\lambda$ values ranged between $-10.0^\circ$ and $10.0^\circ$.  

For the mid-latitude simulations, the latitude of each starspot was selected from a Gaussian distribution centred on one of two regions, namely $45^\circ\pm15^\circ$ and $135^\circ\pm15^\circ$ with the $15^\circ$ limit set at 2-$\sigma$ deviations.

%--------------------------------------------------------------------
\subsection{Simulation execution}\label{Sec:2.3}

A total of $3.6\times10^9$ simulations were conducted and covered 36 different scenarios. Each scenario used a uniform, a polar-biased, or a mid-latitude starspot distribution and one of the starspot filling factors for each star, giving 18 scenarios using a K dwarf and 18 scenarios using an M dwarf (Table\,\ref{Tab.1}). While conducting tests on the simulations, it was found that a variance was detected between the results. Therefore, it was decided that 1000 simulation runs would be conducted for each scenario, where each simulation run would be composed of $10^5$ simulations, giving $10^8$ simulations for each scenario. The final result for each scenario was the mean result from the 1000 simulation runs combined with the 1-$\sigma$ uncertainty.

First, a stellar radius was chosen from a uniform distribution ranging from 0.155\Rsun\ to 0.493\Rsun\ for the M dwarf simulations and 0.568\Rsun\ to 0.761\Rsun\ for the K dwarf simulations. For the simulations that used a polar-biased or mid-latitude starspot distribution, $\lambda$ was selected and applied to the stellar disc. The starspot filling factor limit was then chosen from a Gaussian distribution (see Table\,\ref{Tab.1}). Starspots were then added one at a time (see Section\,\ref{Sec:2.2}) until the stellar disc reached the selected starspot filling factor. The filling factor of each starspot can be found using the solid angle ($\Omega$) of the starspot whereby

\begin{equation}
\Omega = 2\pi R_\mathrm{*}^2\left(1 - \cos r_\mathrm{spot}\right) \ ,
\end{equation}

\noindent where if $r_\mathrm{spot} = 90^\circ$ then the filling factor equates to $2\pi R_\mathrm{*}^2$, half the stellar surface, or a full hemisphere.

Once the starspot filling factor exceeded the selected starspot filling factor limit, the generation of starspots was halted and the $r_\mathrm{spot}$ of the last generated starspot was reduced to bring the simulation starspot filling factor into agreement with the prelected starspot filling factor limit.

The latitude of each starspot was then compared to the latitude of the transit cord. The transit cord was determined by selecting the impact parameter from a uniform distribution. The upper and lower boundary of the transit cord was then calculated by applying $R_\mathrm{p}$ which was drawn from the pre-built distribution (Section\,\ref{Sec:2.2}). This meant that a starspot would be found within the transit cord if $\phi = b\pm R_\mathrm{p}$.

Once the number of starspots within the transit cord had been determined, it was then necessary to determine whether or not the starspots would generate an anomaly detectable in the TESS light curves. This was accomplished by comparing $\Delta F_\mathrm{spot}$ with $k$ (Eq.\,\ref{eq.4}\,\&\,\ref{eq.6}). $\Delta F_\mathrm{spot}$ can be calculated using $r_\mathrm{spot}$ and $\rho_\mathrm{spot}$ \citep{Jeremy2019}:

\begin{equation}
 \Delta F_\mathrm{spot} = \left(1 - \cos r_\mathrm{spot}\right)\left(1 - \rho_\mathrm{spot}\right) \ .
\end{equation}

Subsequently, $\Delta F_\mathrm{spot}$ was compared to the rms scatter of the light curve. If at least one $\Delta F_\mathrm{spot}$ within the transit cord was equal to or greater than 1.5 times the rms scatter of the light curve and satisfied Eq.\,\ref{eq.4}, then the starspot was deemed detectable.

%--------------------------------------------------------------------
\section{Results}\label{Sec:3}

The results from the simulations using a uniform, polar-biased, or mid-latitude starspot distribution are given in Tables\,\ref{Tab.4}\,,\,\ref{Tab.5},\,and\,\ref{Tab.6} respectively. The results indicate that the uniform starspot distributions gave the highest overall starspot detection rates, namely $17.26\%\pm0.48\%$ (for M dwarfs) and $10.66\%\pm0.41\%$ (for k dwarfs). The polar-biased starspot distributions gave the lowest overall starspot detection rates for M and K dwarfs, of $9.95\%\pm0.44\%$ and $5.53\%\pm0.31\%$, respectively. Lastly, the mid-latitude starspot distribution simulations gave overall detection rates of $12.59\%\pm0.47\%$ (for M dwarfs) and $9.59\%\pm0.42\%$ (for k dwarfs).

%--------------------------------------------------------------------
\subsection{Uniform distribution results}\label{Sec:3.1}

The results show that the starspot detection rate reduces with the starspot filling factor (Table\,\ref{Tab.4}), which is expected.

\begin{table} \centering
\caption{\label{Tab.4} Uniform starspot distribution results. The weighted mean of the starspot detection rate of the different filling factors is given in bold.}
\setlength{\tabcolsep}{12pt} \vspace{-5pt}
\begin{tabular}{lcc} 
\hline\hline
 Spectral  & Starspot Filling & Starspot Detection \\   
Type  & Factor (\%) & Rate (\%) \\
\hline
M dwarf & $40.0\%\pm10.0\%$  & $67.0\%\pm1.5\%$ \\ 
M dwarf & $20.0\%\pm5.0\%$  & $46.4\%\pm1.7\%$ \\ 
M dwarf & $4.0\%\pm1.0\%$  & $17.1\%\pm1.2\%$ \\ 
M dwarf & $2.0\%\pm0.5\%$ & $13.2\%\pm1.1\%$\\ 
M dwarf & $0.4\%\pm0.1\%$  & $2.9\%\pm1.0\%$ \\ 
M dwarf & $0.2\%\pm0.05\%$  & $2.5\%\pm1.0\%$ \\ 
\hline
                                    & & {\bf$\mathbf{17.26}$\%$\mathbf{\pm0.48}$\% } \\
\hline
K dwarf & $30.0\%\pm10.0\%$ & $47.9\%\pm1.6\%$ \\
K dwarf & $15.0\%\pm5.0\%$  & $30.8\%\pm1.4\%$ \\
K dwarf & $3.0\%\pm1.0\%$  & $11.6\%\pm1.0\%$\\
K dwarf & $1.5\%\pm0.5\%$  & $9.3\%\pm0.9\%$ \\
K dwarf & $0.3\%\pm0.1\%$ & $2.6\%\pm0.9\%$\\
K dwarf & $0.15\%\pm0.05\%$  & $2.1\%\pm0.8\%$ \\
\hline 
                                    & & {\bf$\mathbf{10.66}$\%$\mathbf{\pm0.41}$\% }  \\
\hline
\end{tabular} 
\end{table}

The simulations indicate that there will be more starspot detections in transit light curves of M dwarf host stars than K dwarf host stars. This is reasonable considering that on average M dwarf stars will have a larger starspot filling factor and because K dwarf stars are larger, $k$ will be smaller; for example, a $13.2\%\pm1.1\%$ detection rate for an M dwarf with a $2.0\%\pm1.0\%$ filling factor and a $11.6\%\pm1.0\%$ detection rate for a K dwarf with a $3.0\%\pm1.0\%$ filling factor (Table\,\ref{Tab.4}). 

\begin{figure*} \centering \includegraphics[width=1.00\textwidth,angle=0]{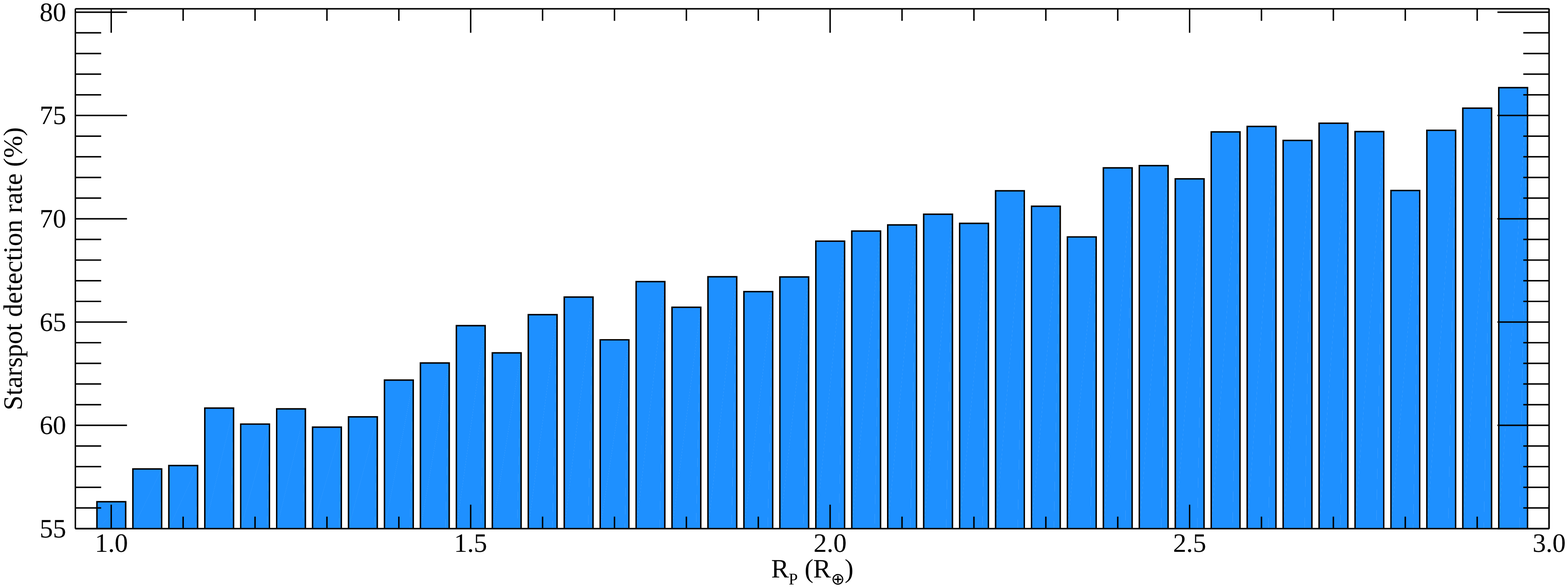}
\caption{\label{fig:1} Starspot detection rate as a function of $R_\mathrm{p}$. From simulation run \#\,{\it540} using a uniform starspot distribution of an M dwarf host star with a starspot filling factor of $40\%\pm10\%$.} \end{figure*}

\begin{figure*} \includegraphics[width=1.00\textwidth,angle=0]{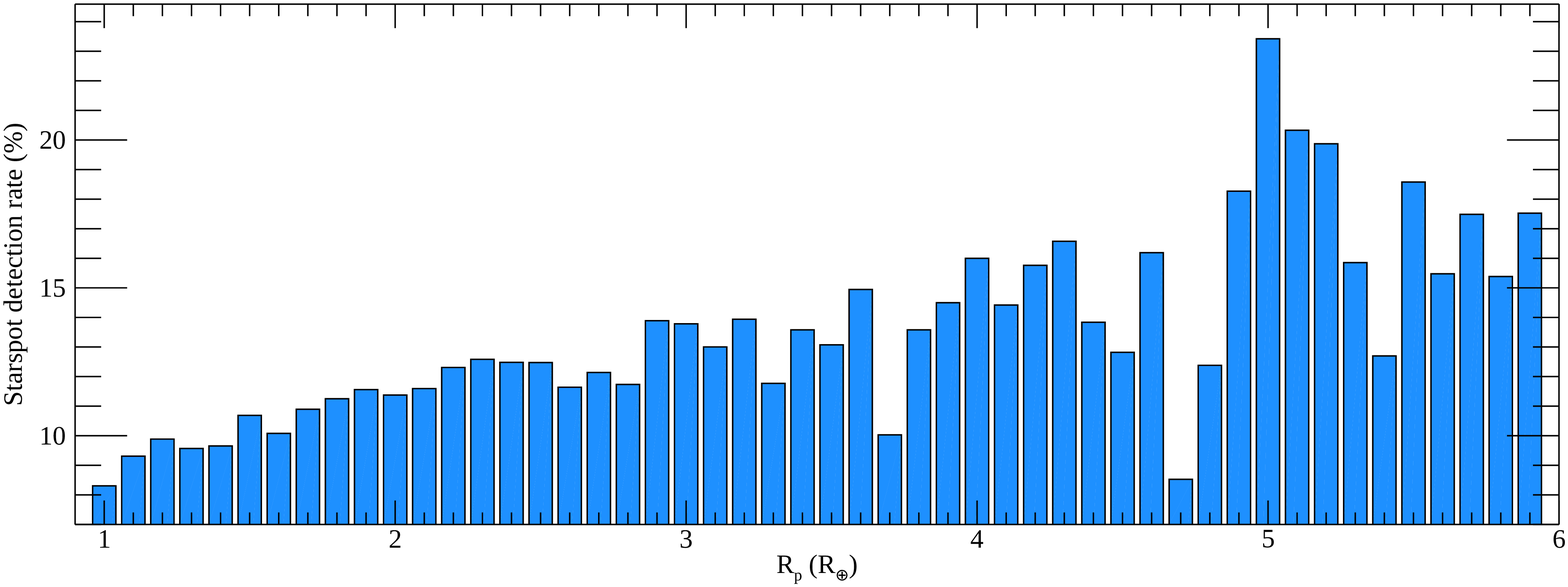} 
\caption{\label{fig:4} Starspot detection rate as a function of $R_\mathrm{p}$. From simulation run \#\,{\it929} using a mid-latitude starspot distribution of a K dwarf host star with a starspot filling factor of $3.0\%\pm1.0\%$.} \end{figure*}

When examining the starspot detection rate as a function of $R_\mathrm{p}$, the results show that the highest detection rates are for the largest planets (see Figs.\,\ref{fig:1} and \ref{fig:4}). When examining the detection rates as a function of $R_\mathrm{p}$ for simulation run \#\,{\it540} of an M dwarf star with a uniform starspot distribution and a filling factor of $40\%\pm10\%$, it can be seen that the detection rate increases from $56\%$ ($R_\mathrm{p} = 1.0\Rearth$) to $76\%$ ($R_\mathrm{p} = 3.0\Rearth$).

%\begin{figure} \includegraphics[width=0.48\textwidth,angle=0]{his2.eps} 
%\caption{\label{fig:2} Number of simulations using values of $k$. From simulation run \#\,{\it118} using a uniform starspot distribution of an M dwarf host star with a starspot filling factor of $40\pm10$\,\%.} \end{figure}

%The simulations drew values of $R_\mathrm{*}$ and $R_\mathrm{p}$ from uniform distributions, however, to achieve a large $k$ requires a small $R_\mathrm{*}$ and a large $R_\mathrm{p}$. In the case of simulation run \#\,{\it118}, this manifested with the mean value of $k$ used in the simulations as $k = 0.063\pm0.029$ which agrees with the mean value of $k$ from the simulations where a starspot was detected, $k = 0.069\pm0.030$ (see Fig.\,\ref{fig:1}).

%\begin{figure} \includegraphics[width=0.48\textwidth,angle=0]{his3.eps} 
%\caption{\label{fig:3} Normalised starspot detection rate as a function of $k$. From simulation run \#\,{\it118} using a uniform starspot distribution of an M dwarf host star with a starspot filling factor of $40\pm10$\,\%.} \end{figure}

%Therefore to determine the starspot detection rate as a function of $k$ requires the number of starspot detections to be normalised against the number of simulations using the corresponding $k$ values. Fig.\,\ref{fig:1} shows the starspot detection rate as a function of $k$. It can be seen that the starspot detection rate increases with increasing $k$, with a starspot detection rate of 30\,\% for $k = 0.019$ and 100\,\% for $k = 0.177$. 

The same trend is seen in the simulations using a K dwarf host star (Fig.\,\ref{fig:4}). For simulation run \#\,{\it929} using a mid-latitude starspot distribution combined with a starspot filling factor of $3.0\%\pm1.0\%$ a starspot detection rate of $8\%$ is found for $R_\mathrm{p} = 1.0\Rearth$ which increases to $17\%$ for $R_\mathrm{p} = 6.0\Rearth$. Though due to the small number of large planets ($>$4\Rearth) used (see Fig\,\ref{fig:1a}), the scatter in the detection rate increases, though it is still possible to see the base trend.

%--------------------------------------------------------------------
\subsection{Polar-biased distribution results}\label{Sec:3.2}

\begin{figure*} \includegraphics[width=0.48\textwidth,angle=0]{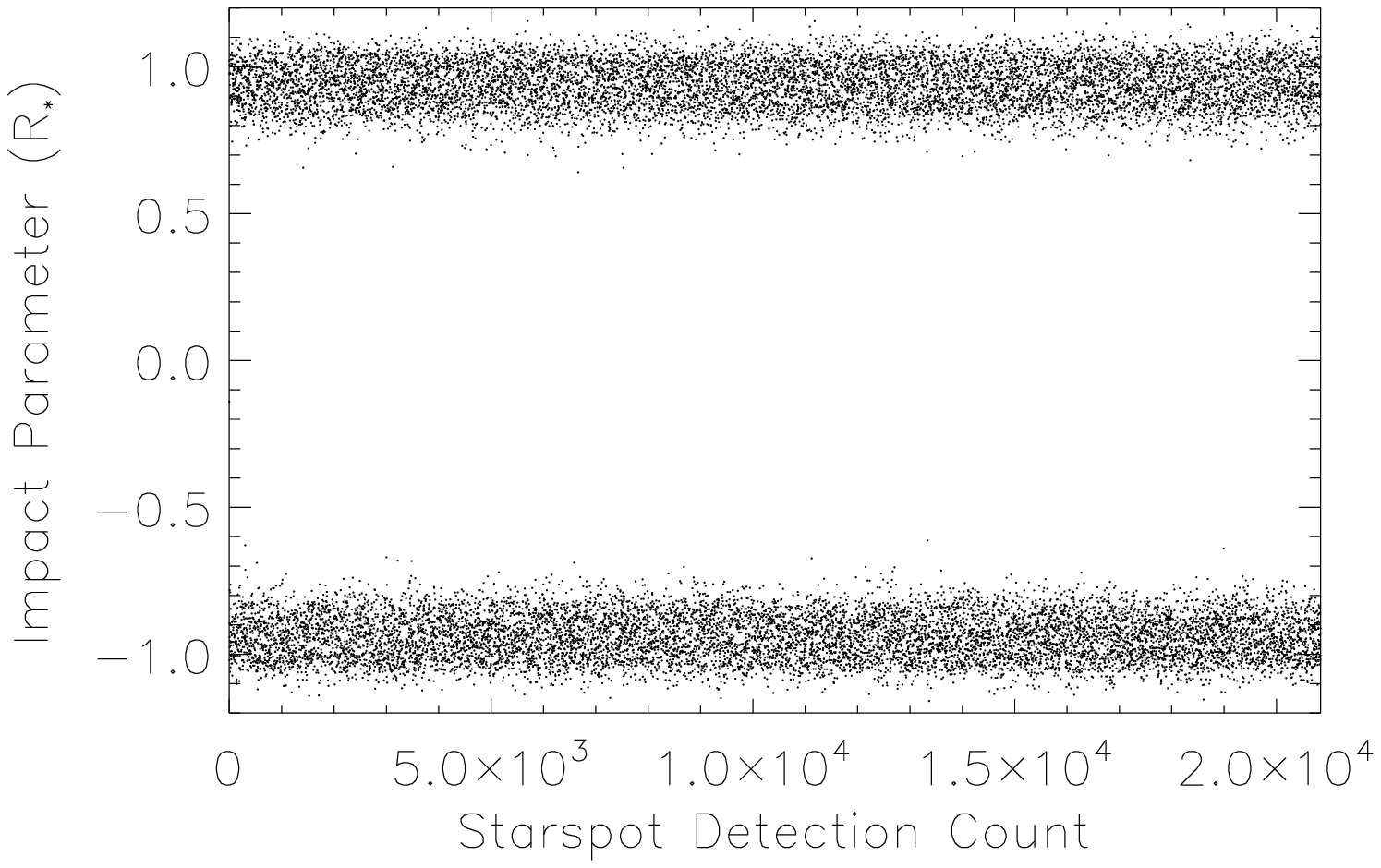} \includegraphics[width=0.48\textwidth,angle=0]{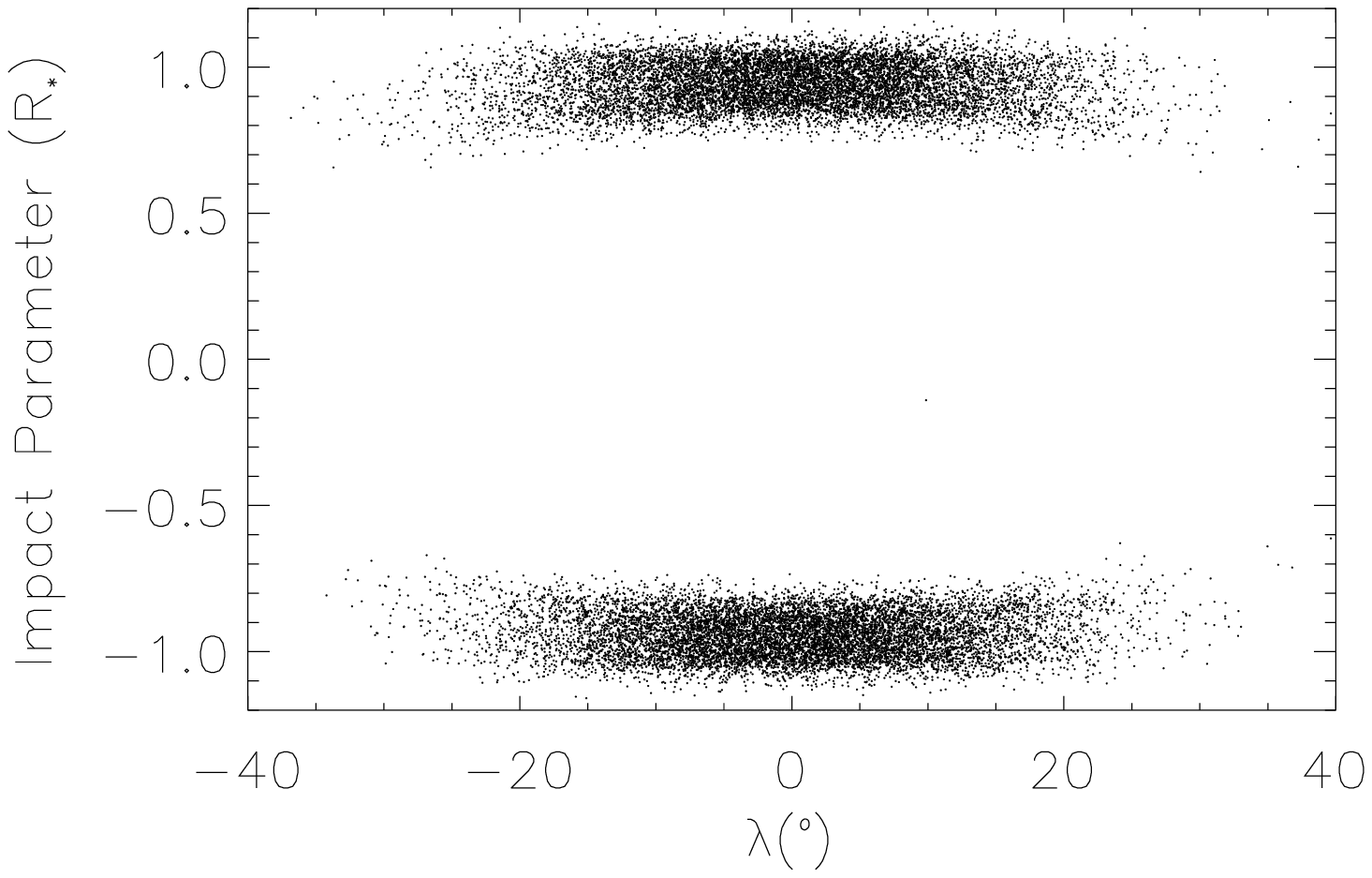}
\caption{\label{fig:5} Impact parameter values in units of $R_\mathrm{*}$ for simulation run \#\,{\it243} using a polar-biased starspot distribution of an M dwarf host star with a starspot filling factor of $20\%\pm10\%$, where starspots were detected. {\it Left:} Impact parameter for simulations with a detected starspot lie close to 1.0\Rsun\ and -1.0\Rsun, indicating the detected starspot was in the polar zone of the star. A few detections have a mid-range impact parameter (e.g. 0.5\Rsun\ and -0.5\Rsun). {\it Right:} Impact parameter as a function of $\lambda$ for simulations with a detected starspot. Here it can be seen that the detections with a mid-range impact parameter took place when $\lambda$ was close to $\pm40^\circ$.} \end{figure*}

The results from a polar-biased starspot distribution (Table\,\ref{Tab.5}) indicate that the starspot detection rate is lower for a polar-biased distribution than for the other distributions. This can be explained when examining the relationship between the detection rate and the impact parameter. For a uniform distribution, the detection rate is independent of the impact parameter, while for the polar biased distribution the detection rate is dependent on the impact parameter (Fig.\,\ref{fig:5}). For a polar distribution of an aligned system, a detection is made when a large impact parameter is selected. Additionally, constraining the surface area where starspots may manifest reduces the detection rates. For instance, the overall detection rate using an M dwarf host star reduces from $17.26\%\pm0.48\%$ for a uniform distribution to $9.95\%\pm0.44\%$ when a polar-biased distribution is used.

\begin{table} \centering
\caption{\label{Tab.5} Polar-biased starspot distribution results. The weighted mean of the starspot detection rate of the different filling factors is given in bold.}
\setlength{\tabcolsep}{12pt} \vspace{-5pt}
\begin{tabular}{lcc} 
\hline\hline
 Spectral  & Starspot Filling & Starspot Detection \\   
Type  & Factor (\%) & Rate (\%) \\
\hline
M dwarf & $40.0\%\pm10.0\%$ & $23.7\%\pm1.4\%$\\ 
M dwarf & $20.0\%\pm5.0\%$  & $21.5\%\pm1.4\%$ \\ 
M dwarf & $4.0\%\pm1.0\%$  & $13.1\%\pm1.1\%$ \\ 
M dwarf & $2.0\%\pm0.5\%$  & $10.8\%\pm1.0\%$ \\ 
M dwarf & $0.4\%\pm0.1\%$  & $3.0\%\pm0.9\%$\\ 
M dwarf & $0.2\%\pm0.05\%$  & $2.9\%\pm0.9\%$ \\ 
\hline       
                                    & & {\bf$\mathbf{9.95}$\%$\mathbf{\pm0.44}$\% } \\
\hline
K dwarf & $30.0\%\pm10.0\%$  & $17.8\pm1.2\%$ \\
K dwarf & $15.0\%\pm5.0\%$ & $15.3\pm1.1\%$ \\
K dwarf & $3.0\%\pm1.0\%$  & $8.8\pm0.9\%$\\
K dwarf & $1.5\%\pm0.5\%$  & $6.5\pm0.8\%$\\
K dwarf & $0.3\%\pm0.1\%$  & $2.0\pm0.7\%$\\
K dwarf & $0.15\%\pm0.05\%$ & $1.9\pm0.5\%$ \\
\hline 
                                    & & {\bf$\mathbf{5.53}$\%$\mathbf{\pm0.31}$\% } \\
\hline 
\end{tabular} 
\end{table}

Examining Fig.\,\ref{fig:5}, we can see that all the simulations where a starspot was detected using a polar-biased distribution had a large impact parameter, $b\approx |1\Rsun|$. Comparing the impact parameter and $\lambda$ for the simulations where a starspot was detected, a strong dependence between the two parameters is seen. Figure\,\ref{fig:5} shows the results from a simulation run (\#\,{\it243}) using a polar-biased starspot distribution of an M dwarf host star with a starspot filling factor of $20\%\pm10\%$. For the majority of simulations when a starspot was detected, the system was aligned ($\lambda = 0^\circ\pm10^\circ$) and the planet transited one of the two stellar polar regions ($b\approx |1\Rsun|$). When $\lambda$ approached either +40$^\circ$ or -40$^\circ$, the simulations indicated a starspot detection when the impact parameter was close to $|0.5\Rsun|$. This dependence is due to the stringent limits placed on and $\lambda$. As discussed in Section\,\ref{Sec:2.2.1} for the polar biased simulations, there was a $95.4\%$ likelihood of a generated starspot lying within 30$^\circ$ of a stellar pole in conjunction with $68.2\%$ of the simulated planetary systems being aligned ($\lambda = 0^\circ\pm10^\circ$).

%--------------------------------------------------------------------
\subsection{Mid-latitude distribution results}\label{Sec:3.2a}

\begin{table} \centering
\caption{\label{Tab.6} Mid-latitude starspot distribution results. The weighted mean of the starspot detection rate of the different filling factors is given in bold.}
\setlength{\tabcolsep}{12pt} \vspace{-5pt}
\begin{tabular}{lcc} 
\hline\hline
Spectral & Starspot Filling & Starspot Detection \\   
Type & Factor (\%) & Rate (\%) \\
\hline
M dwarf & $40.0\%\pm10.0\%$ & $37.3\%\pm1.5\%$ \\ 
M dwarf & $20.0\%\pm5.0\%$ & $31.2\%\pm1.5\%$ \\ 
M dwarf & $4.0\%\pm1.0\%$ & $16.4\%\pm1.2\%$ \\ 
M dwarf & $2.0\%\pm0.5\%$  & $13.1\%\pm1.1\%$ \\ 
M dwarf & $0.4\%\pm0.1\%$  & $2.6\%\pm1.0\%$ \\ 
M dwarf & $0.2\%\pm0.05\%$  & $1.8\%\pm1.0\%$\ \\ 
\hline 
                                    & & {\bf$\mathbf{12.59}$\%$\mathbf{\pm0.47}$\% } \\
\hline
K dwarf & $30.0\%\pm10.0\%$  & $29.7\%\pm1.4\%$ \\
K dwarf & $15.0\%\pm5.0\%$  & $23.3\%\pm1.4\%$ \\
K dwarf & $3.0\%\pm1.0\%$  & $11.5\%\pm1.0\%$\\
K dwarf & $1.5\%\pm0.5\%$  & $9.5\%\pm0.9\%$ \\
K dwarf & $0.3\%\pm0.1\%$  & $2.4\%\pm0.9\%$\\
K dwarf & $0.15\%\pm0.05\%$  & $1.5\%\pm0.9\%$ \\
\hline 
                                    & & {\bf$\mathbf{9.59}$\%$\mathbf{\pm0.42}$\% } \\
\hline 
\end{tabular} 
\end{table}

\begin{figure*} \includegraphics[width=0.48\textwidth,angle=0]{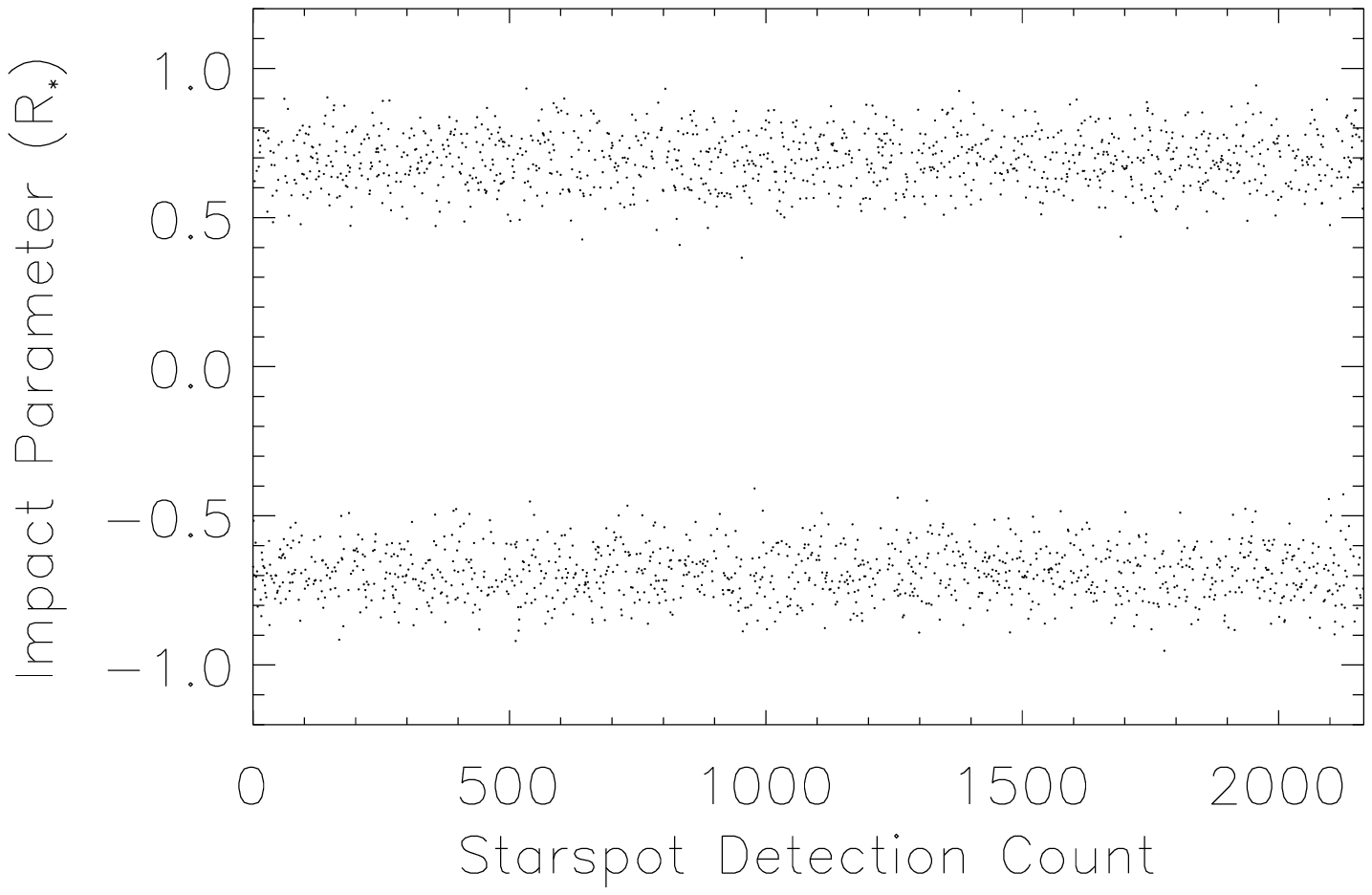} \includegraphics[width=0.48\textwidth,angle=0]{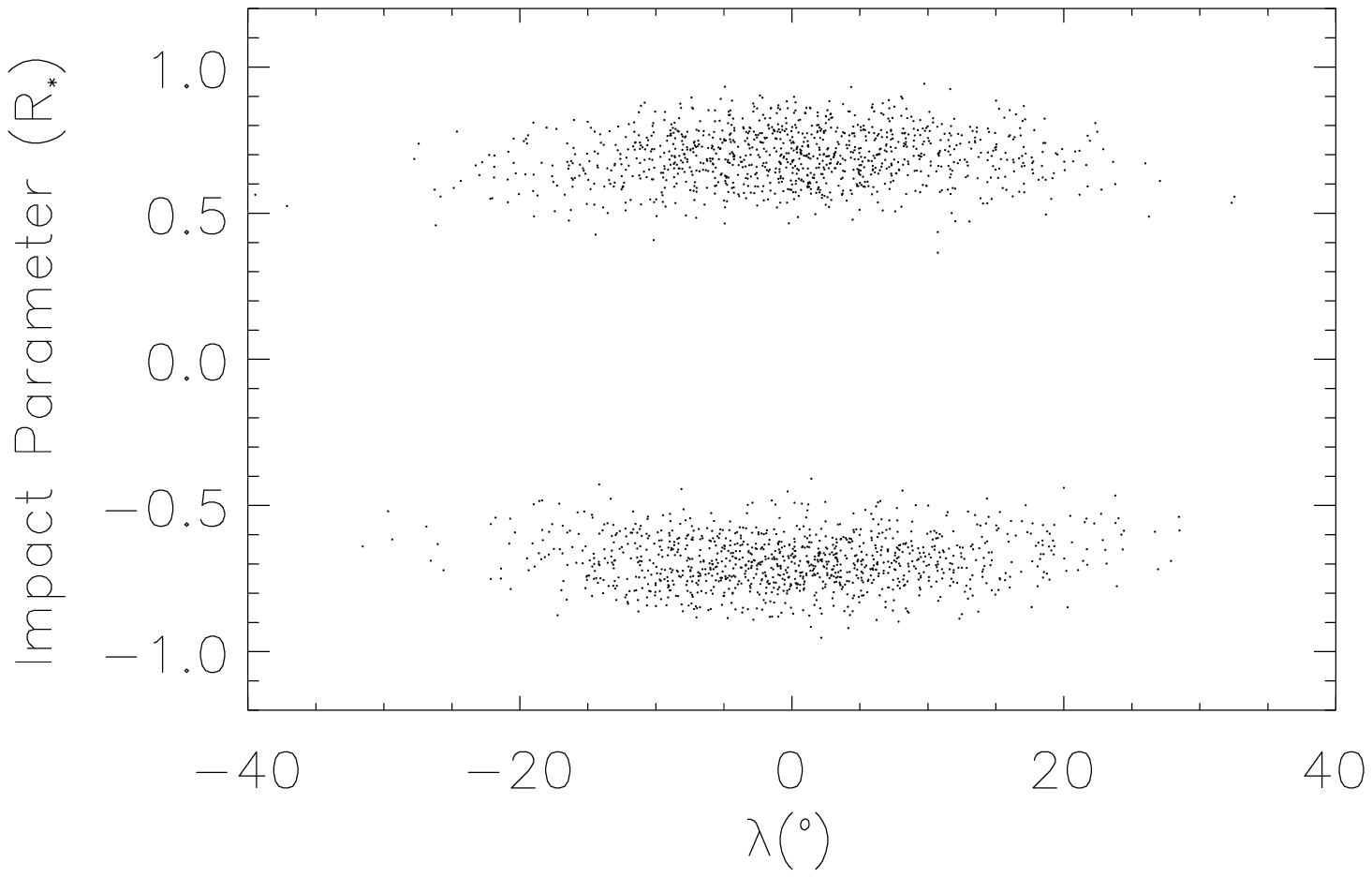}
\caption{\label{fig:5.5} Impact parameter values in units of $R_\mathrm{*}$ for simulation run \#\,{\it63} using a mid-latitude starspot distribution of an K dwarf host star with a starspot filling factor of $0.30\%\pm0.15\%$, where starspots were detected. {\it Left:} Impact parameter for simulations with a detected starspot lie close to 0.7\Rsun\ and -0.7\Rsun\ (stellar latitude 45$^\circ$ and 135$^\circ$, respectively), indicating the detected starspot was in the mid-latitude zone of the star. A few detections had a mid-range impact parameter (e.g. 0.5\Rsun\ and -0.5\Rsun). {\it Right:} Impact parameter as a function of $\lambda$ for simulations with a detected starspot. Here it can seen that the detections with a mid-range impact parameter took place when $\lambda$ was larger than $\pm30^\circ$.} \end{figure*}

The results from the mid-latitude starspot distributions (Table\,\ref{Tab.6}) give a larger detection rate than the polar-biased distribution but a lower rate than the uniform distribution. In the mid-latitude simulations, the surface area of the visible stellar disc for which the starspots are constrained is larger than for the polar-biased distribution, though smaller than for a uniform distribution.

Figure\,\ref{fig:5.5} shows the detection rate results for simulation run (\#\,{\it63}) of a K dwarf host star with a $0.30\%\pm0.15\%$ starspot filling factor using a mid-latitude starspot distribution. The impact parameter of the simulations where a starspot was detected show a consistent value around $|0.7\Rsun|$, which indicates that a starspot was detected when it had a latitude close to 45$^\circ$ or 135$^\circ$. For the few simulations where the impact parameter was $<|0.5\Rsun|$ we can see that $\lambda$ was either +30$^\circ$ or -30$^\circ$. 

%--------------------------------------------------------------------
\subsection{Predictive yields for TESS}\label{Sec:3.2}

\citet{Barclay2018} revised the predicted planet yields of TESS. This latter study showed that TESS is expected to find $1250\pm70$ (within $90\%$ confidence) transiting exoplanets from the 200,000 preselected stars (monitored at 2\,min cadence), with $371\pm38$ expected to be found around M dwarfs and $216\pm30$ around K dwarfs \citep{Barclay2018}.

%We downloaded the catalogue of simulated TESS detections \reff{\citep{Barclay2018}} from the CDS\footnote{Via anonymous ftp from \href{http://cdsarc.u-strasbg.fr/viz-bin/cat?J/ApJS/239/2}{\reff{http://cdsarc.u-strasbg.fr/viz-bin/cat?J/ApJS/239/2}}.}. Assuming the radii limits of M and K dwarfs given in Table\,\ref{Tab.2} (M dwarf: $R_\mathrm{*} =$ 0.155--0.493; K dwarf: $R_\mathrm{*} =$ 0.568--0.761) we determine that TESS should detect $313\pm18$ and $223\pm15$ transiting planets around M and K dwarf stars respectively from the 200,000 preselected stars. We took the same approach as \citet{Sullivan2015} regarding using a Poisson distribution to estimate the uncertainties.

The results given in Tables\,\ref{Tab.4},\,\ref{Tab.5},\,and\,\ref{Tab.6} were then applied to the predicted planetary yields given by \citet{Barclay2018} in order to forecast the frequency of starspot anomalies appearing in TESS exoplanetary transit light curves at two-minute cadence. Table\,\ref{Tab.7} gives the different forecasts for the various starspot filling factors and distributions for M and K dwarf systems observed by TESS. The uncertainties in the final column are dominated by the planetary yield uncertainties.

\begin{table} \centering
\caption{\label{Tab.7} Number of M and K dwarf exoplanetary transit light curves observed by TESS  forecasted to contain observable starspot anomalies. These combine the results presented in Tables\,\ref{Tab.4},\,\ref{Tab.5},\,and\,\ref{Tab.6} with the results from \citet{Barclay2018}. The weighted mean of the forecasted number of transits of the different starspot filling factors is given in bold.} 
\setlength{\tabcolsep}{6pt} \vspace{-5pt}
\begin{tabular}{lcc} 
\hline\hline
 Spectral  & Starspot Filling & Forecasted TESS light\\   
Type   &Factor (\%) & curves containing starspots \\
\hline
\multicolumn{3}{c}{Uniform Distribution} \\
\hline
M dwarf  & $40\%\pm10\%$      & $249\pm32$\\ 
M dwarf  & $20\%\pm5\%$       & $172\pm25$ \\ 
M dwarf  & $4.0\%\pm1.0\%$    & $63\pm11$ \\ 
M dwarf  & $2.0\%\pm0.5\%$    & $49\pm10$ \\ 
M dwarf  & $0.4\%\pm0.1\%$    & $11\pm5$ \\ 
M dwarf  & $0.20\%\pm0.05\%$  & $9\pm5$ \\
\hline
                                    & & $\mathbf{64\pm9}$ \\
\hline
K dwarf  & $30\%\pm10\%$ & $103\pm18$\\
K dwarf  & $15\%\pm5\%$ & $67\pm13$ \\
K dwarf  & $3.0\%\pm1.0\%$ & $25\pm6$\\
K dwarf  & $1.5\%\pm0.5\%$ & $20\pm5$ \\
K dwarf  & $0.3\%\pm0.1\%$ & $6\pm3$\\
K dwarf  & $0.15\%\pm0.05\%$ & $5\pm3$ \\
\hline
                                    & & $\mathbf{23\pm4}$ \\
\hline
\multicolumn{3}{c}{Polar Distribution} \\
\hline
M dwarf  & $40\%\pm10\%$      & $88\pm15$\\ 
M dwarf  & $20\%\pm5\%$       & $80\pm14$ \\ 
M dwarf  & $4.0\%\pm1.0\%$    & $58\pm9$ \\ 
M dwarf  & $2.0\%\pm0.5\%$    & $40\pm8$ \\ 
M dwarf  & $0.4\%\pm0.1\%$    & $11\pm5$ \\ 
M dwarf  & $0.20\%\pm0.05\%$  & $11\pm5$ \\
\hline
                                    & & $\mathbf{37\pm6}$ \\
\hline
K dwarf  & $30\%\pm10\%$ & $47\pm8$\\
K dwarf  & $15\%\pm5\%$ & $33\pm7$ \\
K dwarf  & $3.0\%\pm1.0\%$ & $19\pm5$\\
K dwarf  & $1.5\%\pm0.5\%$ & $14\pm4$ \\
K dwarf  & $0.3\%\pm0.1\%$ & $4\pm2$\\
K dwarf  & $0.15\%\pm0.05\%$ & $4\pm2$ \\
\hline
                                    & & $\mathbf{12\pm2}$ \\
\hline
\multicolumn{3}{c}{Mid-latitude Distribution} \\
\hline
M dwarf  & $40\%\pm10\%$      & $138\pm20$\\ 
M dwarf  & $20\%\pm5\%$       & $116\pm18$ \\ 
M dwarf  & $4.0\%\pm1.0\%$    & $61\pm11$ \\ 
M dwarf  & $2.0\%\pm0.5\%$    & $49\pm9$ \\ 
M dwarf  & $0.4\%\pm0.1\%$    & $10\pm5$ \\ 
M dwarf  & $0.20\%\pm0.05\%$  & $7\pm5$ \\
\hline
                                    & & $\mathbf{47\pm7}$ \\
\hline
K dwarf  & $30\%\pm10\%$ & $64\pm12$\\
K dwarf  & $15\%\pm5\%$ & $50\pm10$ \\
K dwarf  & $3.0\%\pm1.0\%$ & $25\pm6$\\
K dwarf  & $1.5\%\pm0.5\%$ & $26\pm5$ \\
K dwarf  & $0.3\%\pm0.1\%$ & $5\pm3$\\
K dwarf  & $0.15\%\pm0.05\%$ & $3\pm2$ \\
\hline
                                    & & $\mathbf{21\pm4}$ \\
\hline

\end{tabular} 
\end{table}

%--------------------------------------------------------------------
\section{Discussion and Conclusions}\label{Sec:4}

The results from the simulations performed in this study show that the starspot detection rate is higher for uniform starspot distributions than for either the polar-biased or mid-latitude starspot distributions. The results indicate that for a uniform starspot distribution, the starspot detection rates for M and K dwarf stars are $17.26\%\pm0.48\%$ and $10.66\%\pm0.41\%$, respectively. Extrapolating to the number of predicted transiting exoplanets that TESS should detect around M and K dwarf stars allows us to forecast the frequency that starspot anomalies will appear in TESS exoplanetary light curves. For a uniform starspot distribution, a total of $64\pm9$ M dwarf and $23\pm4$ K dwarf TESS exoplanetary light curves are forecasted to contain a starspot anomaly (Table\,\ref{Tab.7}). For a polar-biased distribution, our results show that $9.95\%\pm0.44\%$ of M dwarf exoplanetary light curves and $5.53\%\pm0.31\%$ of K dwarf exoplanetary light curves will show a starspot anomaly. When extrapolating to the number of predicted transiting exoplanets that TESS should detect around M and K dwarf stars, we find a total of $37\pm6$ M dwarf and $12\pm2$ K dwarf TESS exoplanetary light curves are forecasted to contain a starspot anomaly (Table\,\ref{Tab.7}). For a mid-latitude starspot distribution, the simulation results indicate that the starspot detection rates for M and K dwarf stars are $12.59\%\pm0.47\%$ and $9.59\%\pm0.42\%$, respectively. These latter figures combined with the results from \citet{Barclay2018} suggest that, for this type of distribution, $47\pm7$ M dwarf and $21\pm4$ K dwarf TESS exoplanetary light curves will contain a starspot anomaly (Table\,\ref{Tab.7}). We note that this is a lower limit based on the two-minute cadence results from \citet{Barclay2018}, which, when the full frame images (FFIs) are factored in, will increase the forecasted numbers. While it is possible to detect starspot anomalies in 30-minute cadence data (e.g. Kepler 30, \citealt{Sanchis2012}), it is not ideal or well suited.  \citet{Jeremy2019} discussed their finding that the cadence has a strong impact on the detection limits of starspot anomalies where the planetary system has a small $k$. However, this affect is mostly evident in the detection limits of the smallest and/or hottest (low contrast, $\rho \approx 1$) starspots. However, the cadence of TESS FFIs has decreased from 30\,min to 10\,min in the extended mission phase\footnote{See \href{https://heasarc.gsfc.nasa.gov/docs/tess/the-tess-extended-mission.html}{The TESS Extended Mission}}, which will help improve the number of starspot detections from the FFI data.  

\begin{figure} \includegraphics[width=0.48\textwidth,angle=0]{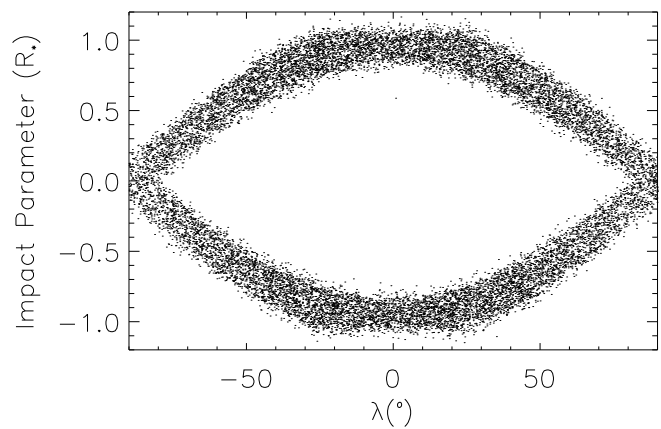} 
\caption{\label{fig:7} Impact parameter values in units of $R_\mathrm{*}$ and $\lambda$ values for simulations where starspots were detected, from a simulation run using a polar-biased starspot distribution of an M dwarf host star with a starspot filling factor of $20\%\pm5\%$. $\lambda$ was selected from a Gaussian distribution of $0\pm75^\circ$.} \end{figure}

Some of the parameters used in the simulations required initial assumptions. For instance in the polar-biased and mid-latitude starspot distributions it was assumed that the distribution of $\lambda$ would follow a similar distribution to that of hot Jupiters around cool stars because of the large convective zones of M and K dwarf stars. The planet radii used in these simulations ranged from 1\,\Rearth\ to 3\,\Rearth\ for the M dwarf simulations and 1\,\Rearth\ to 6\,\Rearth\ for the K dwarf simulations. This size range incorporates both terrestrial (i.e. rocky) and sub-Neptune-class planets. The orbital migration pathways followed by these two classes of planets are unknown and therefore the distribution of $\lambda$ remains unclear. However, the simulations show that the detection rate is not dependent on $\lambda$ but heavily dependent on the latitudinal restriction of the starspots. When separate simulations (using a $20\%\pm5\%$ starspot filling factor of an M dwarf) were run with a relaxed 1-$\sigma$ limit ($\pm75^\circ$) for $\lambda,$ the total number of detections, $21.1\%\pm1.3\%,$ agreed with those from the main simulations within their 1-$\sigma$ uncertainties (see Table\,\ref{Tab.5}). Figure\,\ref{fig:7} shows the results from selecting $\lambda$ from a Gaussian distribution of $0\pm75^\circ$ using a polar-biased starspot distribution of an M dwarf host star with a starspot filling factor of $20\%\pm5\%$. When inspecting Figs.\,\ref{fig:5}\,and\,\ref{fig:7}, it can be seen that the distribution of detections are more uniformly distributed over the $\lambda$ range. This implies that the stellar surface restriction imposed by the polar-biased and mid-latitude distributions has a greater influence on the detection rate than $\lambda$.

None of the published TESS exoplanet transit curves of M and K dwarf host stars are known to contain a starspot anomaly, though three appear to show suspect anomalies (LP\,791-18: \citealt{Crossfield2019}; L\,98-59: \citealt{Kostov2019}; HD-21749: \citealt{Dragomir2019}) and are the subject of further analysis with a dedicated transit-starspot model (e.g. \prism\ \citealt{Jeremy2012,Jeremy2015}). At present\,\footnote{See \href{https://exoplanetarchive.ipac.caltech.edu/}{Exoplanet Archive} accessed 2021/03/01.} there are 17 confirmed M dwarf and 10 K dwarf planetary systems detected by TESS. Using the weighted mean of starspot detections for the uniform starspot distribution,  three\footnote{We assume that multi-planet systems are coplanar.} M dwarf targets and one K dwarf are expected to show starspot anomalies. On the other hand, for a polar-biased distribution, two of the M dwarf systems are expected to show starspot anomalies. Using the mid-latitude distribution results, two M dwarf and one K dwarf planetary system are expected to contain starspot anomalies. However, with the small sample currently available  it is not possible to draw firm conclusions.

Intuitively, for all starspot distributions and spectral classes, the reduction in the starspot filling factor reduces the detection rate. When examining the detection rates of the smallest starspot filling factors, the results from all three distributions are in agreement within their 1-$\sigma$ uncertainties. At these low starspot filling factors, the threshold will be met by the generation of a handful of starspots, each either emulating a small ($r_\mathrm{spot} \approx 1^\circ$) starspot or a group of tiny ($r_\mathrm{spot} < 1^\circ$) starspots (active zone). In this case, it is inconsequential as to whether the small number of starspots is constrained to a certain latitude or not, as the number of latitudes which contain a starspot on the stellar disc in all three distributions will be approximately the same. 

The results from these simulations begin to provide an insight into the activity--age relationship of M and K dwarf stars. Once the TESS data have been fully collected and analysed it will be possible to begin attributing different starspot detection rates to M and K dwarf age ranges. By doing so, it will be possible to map starspot filling factors to different ages and start an investigation into how M and K dwarf activity levels vary with age.

%%%%%%%%%%% https://exoplanetarchive.ipac.caltech.edu/cgi-bin/TblView/nph-tblView?app=ExoTbls&config=planets&constraint=pl_facility+like+%27%25TESS%25%27 %%%%%%%%%%

\begin{acknowledgements}
      We would like to thank the anonymous referee for their helpful comments, which improved the quality of this manuscript. This work was supported by a CONICYT / FONDECYT Postdoctoral research grant, project number: 3180071. JTR thanks the Centro de Astronom\'{i}a (CITEVA), Universidad de Antofagasta for hosting the CONICYT / FONDECYT 2018 Postdoctoral research grant. The following internet-based resources were used in the research for this paper: the NASA Astrophysics Data System; the SIMBAD database and VizieR catalogue access tool operated at CDS, Strasbourg, France; and the ar$\chi$iv scientific paper preprint service operated by Cornell University.
\end{acknowledgements}

\end{document}